\documentclass{mn2e}
\usepackage{times,graphicx,astrobib,amssymb,aas_macros,psfrag}

\def\n{{\bf{\hat{n}}}}
\def\be{\begin{equation}}
\def\e{\end{equation}}
\def\bear{\begin{eqnarray}}
\def\ear{\end{eqnarray}}

\def\begm{\begin{pmatrix}}
\def\enm{\end{pmatrix}}

\def\matrixsymbol{\sf}

\def\apjl{ApJL}

\begin{document}

\title[Constraining large scale HI bias using redshifted 21-cm signal from  the post-reionization epoch]
{Constraining large scale HI bias using redshifted 21-cm signal from  the post-reionization epoch} 

\author[Guha Sarkar, Mitra, Majumdar \& Choudhury ]
{Tapomoy Guha Sarkar$^{1}$\thanks{E-mail: tapomoy@hri.res.in}, 
Sourav Mitra$^{1}$\thanks{E-mail: smitra@hri.res.in},
Suman Majumdar$^{2, 3}$\thanks{E-mail: sumanm@phy.iitkgp.ernet.in} \and
Tirthankar Roy Choudhury$^{1}$\thanks{E-mail: tirth@hri.res.in} \\
$^1$Harish-Chandra Research Institute, Chhatnag Road, Jhusi, Allahabad, 211019 India. \\
$^2$Department of Physics \& Meteorology, IIT, Kharagpur 721302, India.\\
$^3$ Centre for Theoretical Studies, IIT, Kharagpur 721302, India.
}

\maketitle

\date{\today}

\begin{abstract}
In the absence of complex astrophysical processes that characterize
the reionization era, the 21-cm emission from neutral hydrogen (HI) in
the post-reionization epoch is believed to be an excellent tracer of
the underlying dark matter distribution. Assuming a background
cosmology, it is modelled through (i) a bias function $b(k,z)$, which
relates HI to the dark matter distribution and (ii) a mean neutral
fraction ($\bar{x}_{\rm HI}$) which sets its amplitude.  In this
paper, we investigate the nature of large scale $\rm HI$ bias.  The
post-reionization HI is modelled using gravity only N-Body simulations
and a suitable prescription for assigning gas to the dark matter
halos. Using the simulated bias as the fiducial model for $\rm HI$
distribution at $z\leq 4$, we have generated a hypothetical data set
for the 21-cm angular power spectrum ($C_{\ell}$) using a noise model
based on parameters of an extended version of the GMRT. The binned
$C_{\ell}$ is assumed to be measured with SNR $\gtrsim 4$ in the range
$400\leq\ell\leq 8000$ at a fiducial redshift $z=2.5$.  We explore the
possibility of constraining $b(k)$ using the Principal Component
Analysis (PCA) on this simulated data. Our analysis shows that in the
range $0.2 < k < 2$ $\rm Mpc^{-1}$, the simulated data set cannot
distinguish between models exhibiting different $k$ dependences,
provided $1\lesssim b(k)\lesssim 2$ which sets the 2-$\sigma$
limits. This justifies the use of linear bias model on large
scales. The largely uncertain $\bar{x}_{\rm HI}$ is treated as a free
parameter resulting in degradation of the bias reconstruction. The
given simulated data is found to constrain the fiducial $\bar{x}_{\rm
  HI}$ with an accuracy of $\sim 4 \%$ (2-$\sigma$ error).  The method
outlined here, could be successfully implemented on future
observational data sets to constrain $b(k,z)$ and $\bar{x}_{\rm HI}$
and thereby enhance our understanding of the low redshift Universe.
\end{abstract}

\begin{keywords}
cosmology: theory -- large-scale structure of Universe -
cosmology: diffuse radiation
\end{keywords}

\section{Introduction}
Following the epoch of reionization ($z \sim 6$), the low density gas
gets completely ionized (\citeNP{becker01,fan06}). However, a small
fraction of neutral hydrogen (HI) survives, and is confined to the
over-dense regions of the IGM. At this redshifts the bulk of the neutral gas is 
contained in clouds with column density greater than $2 \times
10^{20}$atoms/$\rm cm^2$. Observations indicate that these regions
can be identified as Damped Ly-$\alpha$ (DLA) systems \cite{dla},
which are self-shielded from further ionization and house $\sim 80\%$
of the HI at $1< z < 4$. In this redshift range the neutral fraction
remains constant with $\Omega_{\rm HI} \sim 0.001$ (\citeNP{xhibar,xhibar1,rao2000,xhibar2}).

The distribution and clustering properties of DLAs suggest that they
are associated with galaxies, which represent highly non-linear matter
over densities \cite{haehnelt00}. These clumped HI regions saturate
the Gunn-Peterson optical depth \cite{gunnpeter} and hence cannot be
probed using Ly-{$\alpha$} absorption. They are, however the dominant
source for the 21-cm radiation.  In the post reionization epoch,
Ly-$\alpha$ scattering and the Wouthuysen-Field coupling
(\citeNP{wouth1,purcell,hirev1}) increases the population of the
hyperfine triplet state of HI. This makes the spin temperature $T_s$
much greater than the CMB temperature $T_{\gamma}$, whereby the 21-cm
radiation is seen in emission (\citeNP{madau97,bharad04,zaldaloeb}).
The 21-cm flux from individual HI clouds is too weak ($< 10\mu \rm
Jy$) for detection in radio observations with existing facilities,
unless the effect of gravitational lensing by intervening matter
enhances the image of the clouds significantly \cite{saini2001}.  The
redshifted 21-cm signal however forms a diffuse background in all
radio observations at $z<6$ (frequencies $>203$ MHz). Several radio
telescopes, like the presently functioning
GMRT\footnote{http://www.gmrt.ncra.tifr.res.in/}, and future
instruments MWA\footnote{http://www.mwatelescope.org/} and
SKA\footnote{http://www.skatelescope.org/} aim to detect this weak
cosmological signal submerged in large astrophysical foregrounds
(\citeNP{fg1,fg2,fg3}).
 
The study of large scale structures in redshift surveys and numerical
simulations reveal that the galaxies (for that matter any non linear
structure) trace the underlying dark matter distribution with a
possible bias (\citeNP{mo96,dekel}).  Associating the post-reionization
HI with dark matter halos implies that the gas traces the underlying
dark matter distribution with a possible bias function ${b}(k) =
{[P_{\rm HI}(k)/ P(k)]}^{1/2}$, where ${P}_{\rm HI}(k)$ and ${P}(k)$
denote the power spectra of HI and dark matter density fluctuations
respectively. This function is believed to quantify the clustering property of
the neutral gas. It is believed that on small scales (below the Jean's
length), the bias is a scale dependent function. However, it is
reasonably scale-independent on large scales \cite{fang}. Further, the
bias depends on the redshift. The use of the post-reionization 21-cm
signal (\citeNP{poreion1,poreion2,poreion3,poreion4,poreion5,poreion6})
as a tracer of dark matter opens up new avenues towards various
cosmological investigations (\citeNP{param1,param2,param3,param4}) and
cross-correlation studies (\citeNP{tgs1,tgs2,tgs5}). The underlying bias
model is crucial while forecasting or interpreting some of these
results.

In this paper we have investigated the nature of HI bias in the
post-reionization epoch. The HI fluctuations are simulated at
redshifts $ z <6$ and HI bias is obtained at various redshifts from
the simulated dark matter and HI power spectra. This is similar to the
earlier work by \citeN{bagla} and \citeN{2010ApJ...718..972M}. The simulated bias function is assumed
to be our fiducial model for HI distribution at low redshifts. We have
studied the feasibility of constraining this fiducial model with
observed data.  Here we have focused on the multi frequency angular
power spectrum (MAPS) \cite{datta1}-- measurable directly from
observed radio data and dependent on the bias model.  Assuming a
standard cosmological model and a known dark matter power spectrum we
have used the Principal Component Analysis (PCA) on simulated MAPS data
for a hypothetical radio-interferometric experiment to put constraints
on the bias model. The method is similar to the one used for power
spectrum estimation using the CMB data 
(\citeNP{efsta99,huholder03,leach06}) and constraining reionization 
(\citeNP{mitra1,mitra2}). Stringent constraints on the bias function with
future data sets would be crucial in modelling the distribution of
neutral gas at low redshifts and justify the use of HI as a tracer of
the underlying dark matter field. This would be useful for both
analytical and numerical work involving the post-reionization HI
distribution.

The paper is organized as follows -- in the next section we discuss
the simulation of HI distribution and the general features of the bias
function.  Following that, we discuss the HI multi-frequency angular
power spectrum (MAPS),  a statistical quantifier directly measurable from
radio-interferometric experiments. Finally we use the principal
component analysis to investigate the possibility of constraining the
bias model with simulated MAPS \cite{datta1} data.

\section{Simulation Results - The Bias Model}
\label{sec:simulation_results}

We have obtained the dark matter distribution using the PM N-body code
developed by \citeN{bharadwaj04}, assuming a fiducial cosmological
model (used throughout the paper) $\Omega_m = 0.2726$,
$\Omega_{\Lambda} = 0.726$, $\Omega_b =0.0456$, $h = 0.705$, $T_{cmb}
=2.728 K$ $ \sigma_8 =0.809$, $n_s =0.96$ (all parameters from WMAP 7
year data (\citeNP{komatsu,jarosik})).  We simulate $608^3$ particles in
$1216^3$ grids with grid spacing $0.1\, \rm Mpc$ in a $121.6\, \rm
Mpc^3$ box. The mass assigned to each dark matter particle is $ {
  m_{\rm part}}=2.12 \times 10^8 M_{\odot} h^{-1}$. The initial
particle distribution and velocity field generated using Zel'dovich
approximation (at $z \sim 25$) are evolved only under gravity.  The
particle position and velocities are then obtained as output at
different redshifts $1.5\leq z \leq 4$ at intervals of $\delta z =0.5$.
We have used the Friends-of-Friends algorithm \cite{davis} to
identify dark matter over-densities as halos, taking linking length
$b = 0.2 $ (in units of mean inter-particle distance). This gives a
reasonably good match with the theoretical halo mass function 
(\citeNP{jenkins,shethtormen}) for masses as small as $= 10 m_{\rm
  part}$. The halo mass function obtained from simulation 
is found to be in excellent agreement with the Sheth-Tormen mass
function in the mass range $10^{9} \leq M \leq 10^{13}$ $h^{-1} M_{\odot}$.

We follow the prescription of \citeN{bagla}, to populate the halos
with neutral hydrogen and thereby identify them as DLAs.  Equation (3)
of \citeN{bagla} relates the virial mass of halos, $M$ with its
circular velocity $v_{\rm circ}$. The neutral gas in halos can self
shield itself from ionizing radiation only if the circular velocity is
above a threshold of $v_{\rm circ} =30 {\rm km/sec}$ at $z\sim3$. This
sets a lower cutoff for the halo mass $M_{{\rm min}}$. Further, halos
are populated with gas in a way, such that the very massive halos do
not contain any HI. An upper cut-off scale to halo mass $M_{\rm max}$
is chosen using $ v_{\rm circ} = 200 {\rm km/sec}$, above which we do
not assign any HI to halos. This is consistent with the observation
that very massive halos do not contain any gas in neutral form
\cite{pontzen08}.  The total neutral gas is then distributed such that the 
mass of the gas assigned is proportional to the mass of the halo
between these two cut-off limits.  We note that there is nothing
canonical about this scheme. However, with the basic physical picture
in the background this is the simplest model. Results obtained using
alternative HI assignment schemes are not expected to be drastically
different \cite{bagla}.

\begin{figure}
\psfrag{k in h Mpc-1}[c][c][1][0]{{\bf{\Large $k\,{\rm(Mpc^{-1} h)}$}}}
\psfrag{k3*p(k)/(2*pi*pi)}[c][c][1][0]{{\bf{\Large $k^3 P(k)/(2 \pi^2)$}}}
\psfrag{z=2.5}[c][c][1][0]{{\bf{\Large$z = 2.5$}}}
\includegraphics[width=.4\textwidth, height=0.32\textheight, angle=-90]{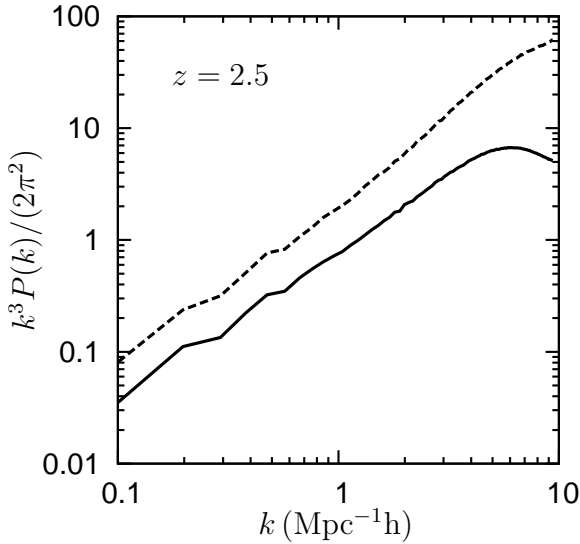}
\caption{The simulated power spectra for dark matter distribution (solid line) 
  and the $\rm HI$ density field (dashed line) at redshift $z=2.5$.}
\label{fig:power}
\end{figure}

\begin{figure}
\psfrag{k in h Mpc-1}[c][c][1][0]{{\bf{\Large $k\,{\rm(Mpc^{-1} h)}$}}}
\psfrag{b(k)}[c][c][1][0]{{\bf{\Large $b^2(k)$}}}
\psfrag{b(z)}[c][c][1][0]{{\bf{\Large$b^2(z)$}}}
\psfrag{z}[c][c][1][0]{{\bf{\Large$z$}}}
\psfrag{k=0.2 h Mpc-1}[c][c][1][0]{{\bf{\large$k = 0.2\,{\rm Mpc^{-1} h}$}}}
\includegraphics[width=.4\textwidth, height=0.32\textheight, angle=-90]{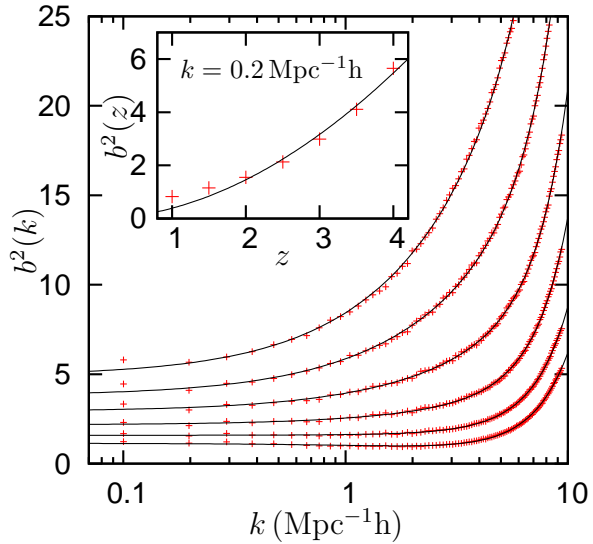}
\caption{The simulated bias function for $z=$1.5, 2.0, 2.5, 3.0, 3.5 
  and 4.0 (bottom to top) showing the scale dependence. The inset shows the 
  variation of the large-scale linear bias as a function of redshift.}
\label{fig:bias}
\end{figure}

Figure \ref{fig:power} shows the simulated power spectra of dark matter
and HI distribution at a fiducial redshift $z=2.5$. The dark matter
power spectrum is seen to be consistent with the transfer function
given by \citeN{hueisen} and the scale invariant primordial power
spectrum (\citeNP{Harri1970,Zeld1972}).  The HI power spectrum has a greater 
amplitude than its dark matter counterpart in the entire $k$-range
allowed by the simulation parameters. Figure \ref{fig:bias} shows the
behaviour of the bias function $b(k,z)$ . We have obtained the scale
dependence of the HI bias for various redshifts in the range $1.5 \leq
z \leq 4$. At these redshifts, the bias is seen to be greater than
unity, a feature that is observed in the clustering of high redshift
galaxies (\citeNP{mo96,wyithe09}). On large cosmological scales the bias remains 
constant and grows monotonically at small scales, where non-linear effects 
are at play. This is a generic feature seen at all redshifts. 
The $k$-range over which the bias function remains scale
independent is larger at the lower redshifts. The linear bias model is
hence seen to hold reasonably well on large scales. The scale
dependence of bias for a given redshift is fitted using a cubic
polynomial with parameters summarized in Table \ref{tab:param}.  The
inset in Figure \ref{fig:bias} shows the redshift dependence of the
linear bias which indicates a monotonic increase. This is also
consistent with the expected $z$-dependence of high redshift galaxy
bias.  The behaviour of the linear bias for small $k$-values as a
function of redshift is non-linear and can be fitted by an approximate
power law of the form $\sim z^{2}$. This scaling relationship of
bias with is found to be sensitive to the mass resolution of the
simulation. The similar dependence of HI bias with $k$ and $z$ has
been observed earlier by \citeN{bagla} with a computationally robust
Tree N-body code. Here we show that, the same generic features and
similar scaling relations for bias can be obtained by using a simpler
and computationally less expensive PM N-body code. Our aim is to use
this scale and redshift dependence of bias, obtained from our
simulation as the fiducial model for the post reionization HI
distribution. We shall subsequently investigate the feasibility of
constraining this model using Principal Component Analysis (PCA) on
simulated MAPS data.


\begin{table}
\centering
\begin{tabular}{c|c|c|c|c}
\hline
\hline
 $z$ & $ c_{3}$ & $ c_{2}$ & $ c_{1}$ & $ c_{0}$ \\
\hline
 1.5 & 0.0029 & 0.0365 & -0.1561 & 1.1402 \\
 2.0 & 0.0052 & 0.0177 & 0.0176 & 1.5837 \\
 2.5 & 0.0101 & -0.0245 & 0.3951 & 2.1672 \\
 3.0 & 0.0160 & -0.0884 & 1.0835 & 2.9287 \\
 3.5 & 0.0234 & -0.1537 & 2.1854 & 3.8050 \\
 4.0 & 0.0248 & -0.1655 & 3.6684 & 4.9061 \\
\hline
\hline
\end{tabular}
\caption{The fit parameters for bias function of the form 
  $b^{2}(k) = c_{3} k^3 +   c_{2}k^2 +  c_{1} k +  c_{0}$ for various 
  redshifts $1.5\leq z\leq 4.0$.}
\label{tab:param}
\end{table}

\section{HI 21-cm angular power spectrum - Simulated data}
\label{sec:simulated_data}
Redshifted 21-cm observations have an unique advantage over other
  cosmological probes since it maps the 3D density field and gives a
  tomographic image of the Universe. In this paper we have
quantified the statistical properties of the fluctuations in the
redshifted 21-cm brightness temperature $ T(\n, z)$ on the sky is
quantified through the multi frequency angular power spectrum MAPS,
defined as $C_{\ell} (\Delta z) = \langle a_{\ell m}(z) a^{*}_{\ell
  m}(z + \Delta z)\rangle$, where $ a_{\ell m}(z) = \int d \Omega_{\n}
Y^{*}_{\ell m} (\n) T(\n, z)$.  This measures the correlation of the
spherical harmonic components of the temperature field at two redshift
slices separated by $\Delta z$.  In the flat-sky approximation and
incorporating the redshift space distortion effect we have
\cite{datta1} \be C_{\ell} = \frac{{\bar T}^2}{\pi r^2}\int
\limits_0^\infty dk_{\parallel} \cos (k_{\parallel}\Delta r) P^s_{\rm
  HI} ({\bf k})
\label{eq:angps}
 \e for correlation between HI at comoving distances $r$ and $ r +
 \Delta r$, $\bar T = 4 \rm mK ( 1 + z )^2 \left( \frac{\Omega_b
   h^2}{0.02} \right ) \frac{H_0}{H(z)}\left( \frac{0.7}{h} \right )
 $, $k = \sqrt { {\left( \frac{\ell}{r}\right)}^2 + k_{\parallel}^2}$
 and $ P^s_{\rm HI}$ denotes the redshift space HI power spectrum
 given by \be P^s_{\rm HI}({\bf k}) ={\bar{x}_{\rm HI}}^2 b^2(k, z)
 D_{+}^2\left[ 1 + \beta {\left ( \frac{k_{\parallel}}{k} \right )}^2
   \right ]^2 P(k) \e where the mean neutral fraction $\bar{x}_{\rm
   HI} $ is assumed to have a fiducial value $2.45 \times 10^{-2}$,
 $\beta = f/b(k,z)$, $f=d\ln D_{+}/d\ln a$ where, $D_{+}$ represents
 the growing mode of density perturbations, $a$ is the cosmological
 scale factor and $P(k)$ denotes the present day matter power
 spectrum.  

  We use MAPS as an alternative to the more commonly used 3D power
  spectrum since it has a few features that makes its measurement more
  convenient.  Firstly we note that as a function of $\ell$ (angular
  scales) and $\Delta z$ (radial separations) the MAPS encapsulate the
  entire three dimensional information regarding the HI distribution.
  In this approach, the fluctuations in the transverse direction are
  Fourier transformed, while the radial direction is kept unchanged  in
  the real frequency space. No cosmological information is however
  lost.  Secondly, 21-cm signal is deeply submerged in astrophysical
  foregrounds. These foregrounds are known to have a smooth and slow
  variation with frequency, whereas the signal is more localized along
  the frequency axis.  The distinct spectral ( $\Delta z$ ) behaviour
  has been proposed to be an useful method to separate the
  cosmological signal from foreground contaminants. Infact it has been
  shown that foregrounds can be completely removed by subtracting out
  a suitable polynomial in $\Delta \nu$ from $C_{\ell}(\Delta \nu)$
  (\citeNP{ghosh2}). It is hence advantageous to use MAPS which
  maintains the difference between the frequency and angular
  information in an observation. The 3D power spectrum on the contrary
  mixes up frequency and transverse information through the full 3D
  Fourier transform.  Further, for a large band width radio
  observation, covering large radial separations light cone effect is
  expected to affect the signal. This can also be easily incorporated
  into MAPS unlike the 3D power spectrum which mixes up the
  information from different time slices.  The key advantage, however,
  in using the angular power spectrum is that it can be obtained
  directly from radio data. The quantity of interest in
  radio-interferometric experiments is the complex Visibility $
  {\mathcal{V}} ( {\bf U} , \nu)$ measured for a pair of antennas
  separated by a distance ${\bf d}$ as a function of baseline ${\bf U}
  = {\bf d}/ \lambda$ and frequency $\nu$. The method of Visibility
  correlation to estimate the angular power spectrum has been well
  established (\citeNP{poreion1,bali}). This follows from the fact
  that $\langle {\mathcal{V}} ( {\bf U} , \nu) {\mathcal{V}} ^{*}(
  {\bf U} , \nu + \Delta \nu) \rangle \propto C_{\ell}(\Delta
  \nu)$. Here the angular multipole $\ell$ is identified with the
  baseline $ U $ as $\ell = 2\pi U$ and one has assumed that the
  antenna primary beam is either de-convolved or is sufficiently
  peaked so that it maybe treated as a Dirac delta function. Further
  the constant of proportionality takes care of the units and depends
  on the various telescope parameters.

The angular power spectrum at a multipole $\ell$ is obtained by
projecting the 3D power spectrum. The integral in Equation
\ref{eq:angps}, sums over the modes whose projection on the plane of
the sky is $\ell/r$. Hence, $C_{\ell}$ has contributions from matter
power spectrum only for $k > \ell/r$.  The shape of $C_{\ell}$ is
dictated by the matter power spectrum $P(k)$ and the bias $b(k)$. The
amplitude depends on quantities dependent on the background
cosmological model as well as the astrophysical properties of the
IGM. We emphasize here that, the mean neutral fraction and the HI bias
are the only two non-cosmological parameters in our model for the HI
distribution at low redshifts. Predicting the nature of $C_{\ell}$ in
a given cosmological paradigm is then crucially dependent on the
underlying bias model and the value of the neutral fraction.

  The $\Delta \nu$ dependance of the MAPS $C_{\ell}(\Delta \nu)$
  measures the correlation between the various 2D modes as a function
  of radial separation $\Delta r$ ($ \Delta \nu$). The signal is seen
  to decorrelate for large radial separations, the decorrelation being
  faster for larger $\ell$ values.  For a given $\ell$, one gets
  independent estimates of $C_{\ell}$ for radial separations greater
  than the correlation length.  Projection of the 3D power spectrum
  leads the availability of fewer Fourier modes. However, for a given
  band width $B$, one may combine the signals emanating from epechs
  separated by the correlation length $\Delta \nu_C$ in the radial
  direction. Noting that the amplitude of the signal does not change
  significantly over the radial separation corresponding to the band
  width, one has $ \sim B/\Delta \nu_c$ independent measurements of
  $C_{\ell}(\Delta z = 0)$. We have adopted the simplified picture
  where the noise in $C_{\ell}(\Delta z = 0)$ gets reduced owing to
  the combination of these $ B/\Delta \nu_c$ realizations.  A more
  complete analysis would incorporate the correlation for $\Delta \nu
  < \Delta \nu_c$. We plan to take this up in a future work.

\begin{figure}
  \includegraphics[width=0.48\textwidth,height=0.34\textheight,
    angle=270]{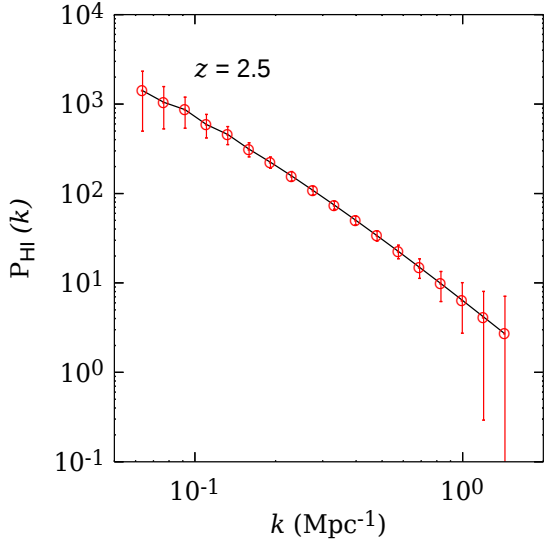}
  \caption{The theoretical 3D $\rm HI$ power spectrum $P^s_{\rm HI}(k)$ 
  for $z=2.5$ as a function of $k$, at $\mu = 0.5$. The points 
  with 2-$\sigma$ error-bars represent the hypothetical binned data.}
\label{fig:threedpk}
\end{figure}

\begin{figure}
\psfrag{3-sigma error}[c][c][1][0]{{\bf{\Large $\sigma$}}}
\psfrag{C_l}[c][c][1][0]{{\bf{\Large $\sigma$}}}
\includegraphics[width=0.48\textwidth,height=0.34\textheight, angle=270]{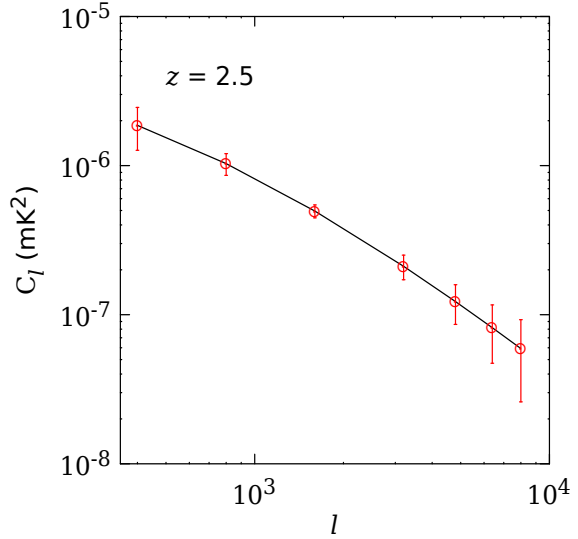}
  \caption{The theoretical  angular power spectrum $C_{\ell}$ 
  for $z=2.5$ as a function of a $\ell$. The points 
  with 2-$\sigma$ error-bars represent the hypothetical data.}
\label{fig:cell}
\end{figure}

Figure \ref{fig:threedpk} shows the 3D HI power spectrum 
at the fiducial redshift $ z = 2.5$ obtained
using the dark matter power spectrum of \citeN{hueisen}. We have used
the WMAP 7 year cosmological model throughout. Figure \ref{fig:cell}
shows the corresponding HI angular power spectrum. 
The shape of $C_{\ell}$ is dictated by the shape of the matter power
spectrum, the bias function, and the background cosmological
model. The amplitude is set by various quantities that depend on the
cosmological model and the growth of linear perturbations. The global
mean neutral fraction also appears in the amplitude and plays a
crucial role in determining the mean level for 21-cm emission.  Hence,
for a fixed cosmological model, the bias and the neutral fraction,
solely determine the fluctuations of the post-reionization HI density
field.  We have used the bias model obtained from numerical
simulations in the last section to evaluate the $C_{\ell}$.  We assume
that the binned angular power spectrum is measured at seven $\ell$
bins $-$ the data being generated using Equation \ref{eq:angps} using
the fiducial bias model.

The noise estimates are presented using the formalism used by
\citeN{param4} for the 3D power spectrum and \citeN{bali} and
\citeN{bagla} for the angular power spectrum.  We have used
hypothetical telescope parameters for these estimates. We consider
radio telescope with 60 GMRT like antennae (diameter 45 m) distributed
randomly over a region $1 \rm km \times 1 \rm km$.  We assume $T_{sys}
\sim 100 K$.  
We consider a  a radio-observation at frequency $ \nu =
405 \rm MHz$ with a bandwidth $B = 32 \rm MHz$ for an observation time
of $1000$ hrs. 

  In order to attain desired sensitivities we have assumed that the
  data is binned whereby several nearby $\ell-$ modes are combined to
  incresase the SNR.  Furthur, in the radial direction, the signal is
  assumed to decorrelate for $\Delta \nu > 0.5 \rm MHz$, so that we
  have $64$ independent measurements of $C_{\ell}$ for the given band
  width of $32 \rm MHz$. The $7$-$\ell$ bins chosen here allows the
  binned power spectrum to be measured at a SNR $\gtrsim 4$ in the
  entire range $400 \leq \ell \leq 8000$.  One would ideally expect to
  measure the power spectrum at a large number of $\ell$ values which
  would necessarily compromise the obtained sensitivities.  With the
  given set of observational parameters, one may, in principle choose
  a finer binning. It shall however degrade the SNR below the level of
  detectability.  Choosing arbitrarily fine $\ell-$ bins and
  simulataneously maintaining the same SNR would require improved
  observational parameters which may be unreasonable if not
  impossible.  The same reasoning applies to noise estimation for the
  3D power spectrum where for a given set of observational parameters,
  the choice of $k-$ bins is dictated by the requirement of
  sensitivity. In the figure \ref{fig:threedpk}, showing the 3D power
  spectrum  a $4-\sigma$ detection of $P_{\rm HI}(k)$ in
  the central bin requires the full $k-$ range to be divided into $18$
  equal logarithmic bins for the same observational parameters.

The noise in $C_{\ell}$ and $P_{\rm HI}(k)$ is dominated by cosmic
variance at small $\ell / k $ (large scales), whereas, instrumental
noise dominates at large $\ell / k$ values (small scales). We point
out that the error estimates predicted for a hypothetical observation
are based on reasonable telescope parameters and future observations
are expected to reflect similar sensitivities.

We note here that several crucial observational difficulties hinder
$C_{\ell}$ to be measured at a high SNR.  Separating the astrophysical
foregrounds, which are several order larger in magnitude than the
signal is a major challenge
(\citeNP{fg1,fg2,fg3,ghosh1,ghosh2}). Several methods have been
suggested for the removal of foregrounds most of which uses the
distinct spectral property of the 21 cm signal as against that of the
foreground contaminants.  The multi frequency angular power spectrum
(MAPS) $C_{\ell}(\Delta \nu)$ is itself useful for this purpose
(\citeNP{ghosh1,ghosh2}). Whereas this signal $C_{\ell}(\Delta \nu)$
decorrelates over large $\Delta \nu$, the foregrounds remain
correlated $-$ a feature that maybe used to separate the two.  In our
subsequent discussions we assume that the foregrounds have been
removed.  As mentioned earlier, the angular power spectrum can
directly be measured from raw visibility data.  One requires to
incorporate the primary beam of the antenna in establishing this
connection \cite{bali}.  In this paper we assume that such
difficulties are overcome and the angular power spectrum is measured
with sufficiently high SNR.

In the next section we use the $C_{\ell}$ data generated with these
assumptions to perform the PCA.  If the 3D HI power spectrum is
  measured at some $(k, \mu)$ it would be possible to determine the
  bias directly from a knowledge of the dark matter power
  spectrum. The bias would be measured at the $k-$ values where the
  data is available. The results for the 3D analysis is summarized in
  section \ref{sec:discussions}.

\section{Principal Component Analysis}
\label{sec:pca_theory}

\begin{figure*}
\psfrag{k}[c][c][1][0]{{\bf{\Large $\tilde{G}$}}}
\includegraphics[width=0.70\textwidth,height=0.405\textheight,angle=0]{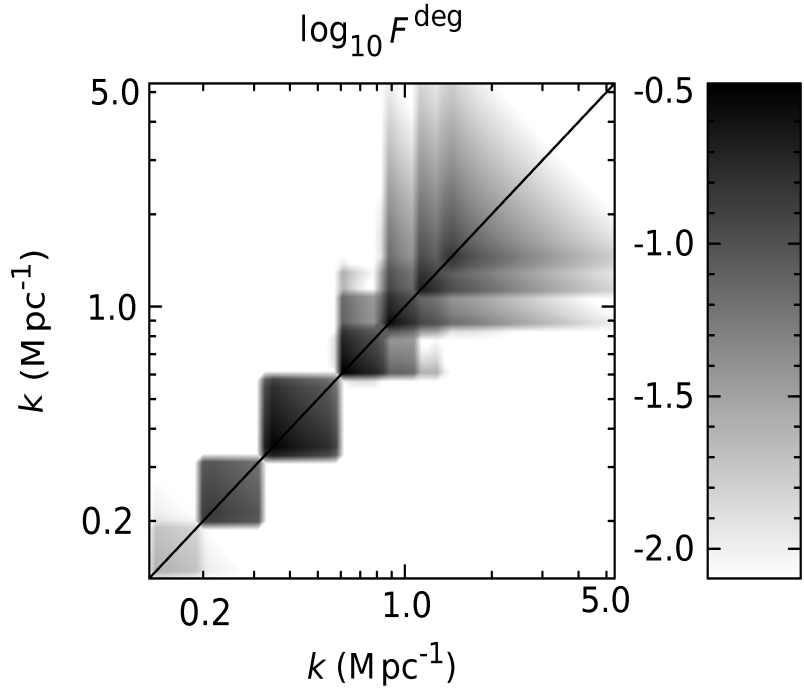}
  \caption{The degraded Fisher matrix $F_{ij}^{deg}$ in the $k-k$ plane.}
\label{fig:fisherplot}
\end{figure*}

In this section, we discuss the principal component method towards
constraining the bias function using $C_{\ell}$ data. We consider a
set of $n_{\rm obs}$ observational data points labeled by ${\cal
  C}_{\ell_{obs}}$ where $\ell_{obs}$ runs over the different $\ell$
values for which $C_{\ell}$ is obtained (Fig. \ref{fig:cell}).

In our attempt to reconstruct $b(k)$ in the range $[k_{\rm min},
  k_{\rm max}]$, we assume that the bias which is an unknown function
of $k$, can be represented by a set of $n_{\rm bin}$ discrete free
parameters $b_{i} = b(k_i)$ where the entire k-range is binned such
that $k_i$ corresponds to the $i^{th}$ bin of width given by 
\be
\Delta \ln k_{i} = \frac{\ln k_{\rm max}- \ln k_{\rm min}}{n_{\rm
    bin}-1} 
\e 
We have chosen $n_{bin} = 61$ and a $k-$range $0.13\leq
k\leq 5.3$ $\rm Mpc^{-1}$. Our choice is dictated by the fact that for
$k<0.13$ $\rm Mpc^{-1}$, the $C_{\ell}$ corresponding to the smallest
$\ell$ is insensitive to $b(k)$ and for $k>5$ $\rm Mpc^{-1}$ there is
no data probing those scales. This truncation is also justified as the
Fisher information matrix, we shall see, tends to zero beyond this
$k-$range.

The Fisher matrix is constructed as
\begin{equation}
 F_{ij}=\sum_{\ell_{obs}} \frac{1}{\sigma_{\ell_{obs}}^2}
\frac{\partial {C}_{\ell_{obs}}^{\rm th}}{\partial b^{\rm fid}(k_i)}
\frac{\partial {C}_{\ell_{obs}}^{\rm th}}{\partial b^{\rm fid}(k_j)},
\label{eq:fisher}
\end{equation}
where ${C}_{\ell_{obs}}^{\rm th}$ is the theoretical
(Eq.\ref{eq:angps}) ${C}_{\ell}$ evaluated at $\ell = \ell_{obs}$
using the fiducial bias model $b^{\rm fid}(k)$ and
$\sigma_{\ell_{obs}}$ is the corresponding observational error.
The data is assumed to be such that the covariance matrix is diagonal whereby only the variance $\sigma_{\ell_{obs}}$ suffices.

 The fiducial model for bias is, in principle, expected to be close to
 the underlying ``true'' model. In this work we have taken $b^{\rm fid}(k)$
 to be the fitted polynomial  obtained in the earlier section which matches the
 simulated bias  up to an acceptable accuracy.

In the model for HI distribution at low redshifts, the mean neutral
fraction crucially sets the amplitude for the power spectrum. However,
a lack of precise knowledge about this quantity makes the overall
amplitude of $C_{\ell}$ largely uncertain. To incorporate this we have
treated the quantity $\bar{x}_{\rm HI}$ as an additional free parameter
over which the Fisher matrix is marginalized.
The corresponding  degraded Fisher matrix is given by
\be {\matrixsymbol F}^{deg} = {\matrixsymbol F} - {\matrixsymbol B}
    {\matrixsymbol F'}^{-1} {\matrixsymbol B}^T 
\label{eq:fishdeg}
\e where ${\matrixsymbol F}$ is the original $ n_{\rm bin} \times
n_{\rm bin}$ Fisher matrix corresponding to the parameters $b_i$,
${\matrixsymbol F'}$ is a $ 1 \times 1 $ Fisher matrix for the
additional parameter  $\bar{x}_{\rm HI}$, and ${\matrixsymbol B}$ is a
$n_{\rm bin} \times 1$-dimensional matrix containing the cross-terms.
We shall henceforth refer to ${\matrixsymbol F}^{deg}$ as the Fisher
matrix and implicitly assume that the marginalization has been
performed.

The Fisher matrix obtained using Eq.\ref{eq:fisher} and
Eq.\ref{eq:fishdeg} is illustrated in Figure \ref{fig:fisherplot} as a 
shaded plot in the $k-k$ plane. The matrix shows a band diagonal 
structure with most of the information accumulated in discrete regions 
especially corresponding to the $k-$modes for which the data is available. 
In the region $k>2$ and $k<0.2$ $\rm Mpc^{-1}$, the value of 
$F_{ij}$ is relatively small, implying that one cannot constrain $b(k)$ 
in those $k-$bins from the data set we have considered in this work.

A suitable choice of basis ensures that the parameters are not correlated.
This amounts to writing the Fisher matrix in its eigen basis.
Once the Fisher matrix is constructed, we  determine its
eigenvalues and corresponding eigenvectors. The
orthonormality and completeness of the eigenfunctions, allows us to  expand
the deviation of $b(k_i)$ from its fiducial model, $\delta
b_{i}=b(k_i)-b^{\rm fid}(k_i)$, as
\begin{equation}
 \delta b_i=\sum_{p=1}^{n_{\rm bin}} m_pS_{p}(k_i)
\label{eq:pcaexp}
\end{equation}
where $S_p(k_i)$ are the principal components of $b(k_i)$ and 
$m_p$ are the suitable expansion coefficients. The advantage is that, unlike $b(k_i)$, 
the coefficients $m_p$ are uncorrelated.

\begin{figure}
  \psfrag{mode i}[c][c][1][0]{{\bf{\Large $k$}}}
  \includegraphics[width=0.43\textwidth,angle=270]{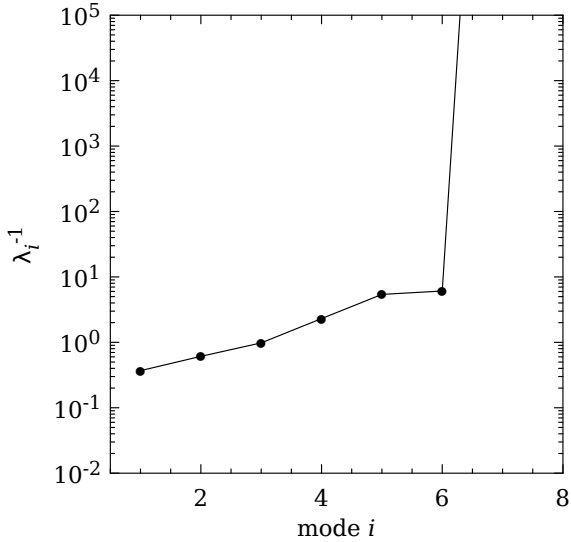} 
  \caption{The inverse of eigenvalues of the degraded Fisher matrix $F_{ij}^{deg}$ which essentially 
   measures the variance on the corresponding coefficient.}
\label{fig:eigenval}
\end{figure}

\begin{figure}
  \includegraphics[height=0.53\textwidth, angle=0]{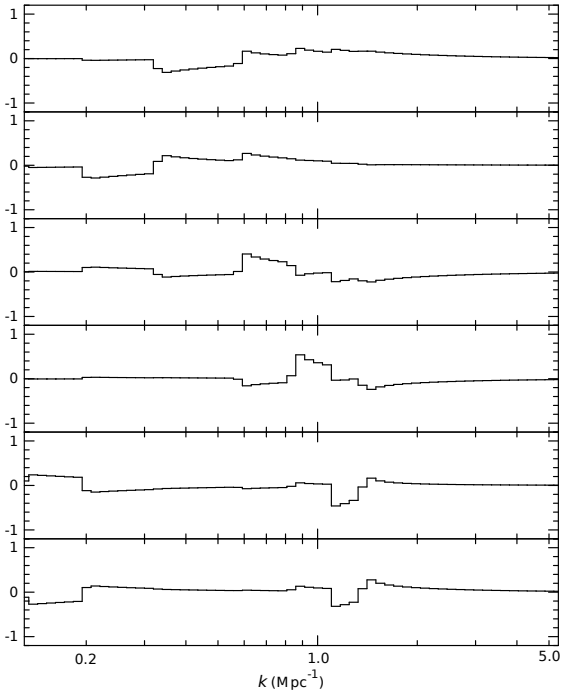}
    \caption{The first 6 eigenmodes of the degraded Fisher matrix.}
\label{fig:eigenvec}
\end{figure}

Figure \ref{fig:eigenval} shows the inverse of the largest
eigenvalues.  Beyond the first six, all the eigenvalues are seen to be
negligibly small. It is known that the largest eigenvalue corresponds 
to minimum variance set by the Cramer-Rao bound and vice versa. 
This implies that the errors in $b(k)$ would increase drastically if 
modes $i > 6$ are included. Hence, most of the relevant information 
is essentially contained in the first six modes with larger eigenvalues. 
These normalized eigenmodes are shown in the Figure \ref{fig:eigenvec}. 
One can see that, all these modes almost tend to vanish for $k>2$ and $k<0.2$ 
$\rm Mpc^{-1}$, as the Fisher matrix is vanishingly small in these 
regions. The positions of the spikes and troughs in these modes are related 
to the presence of data points and their amplitudes depend on the 
corresponding error-bars (smaller the error, larger the amplitude).

  The fiducial model adopted in our analysis may be different from
  the true model which dictates the data. Clearly, the reconstruction
  would be poor for wide discrepancies between the two. In our
  analysis, the simulated bias serves as the input.  In the absence of
  many alternative models for large scale HI bias, this serves as a
  reasonable fiducial model.

\begin{figure}
\includegraphics[width=0.48\textwidth,height=0.34\textheight, angle=0]{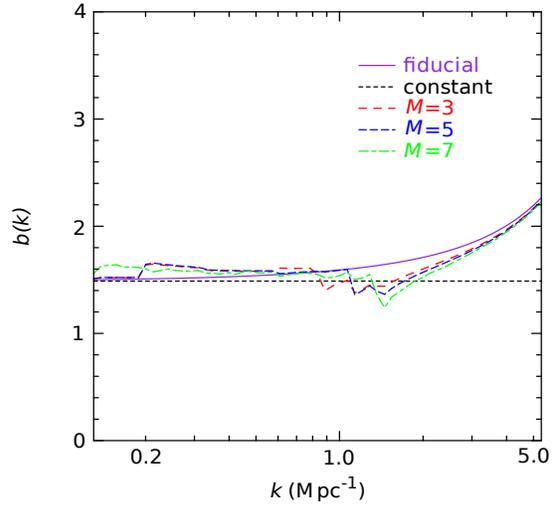}
  \caption{The fiducial and constant (true) bias models are shown. The
    reconstruction of the true model is shown for cases where number
    of PCA modes considered are $M = 3, 5, 7$}
\label{fig:recon}
\end{figure}

We assume that one can then reconstruct the function $\delta b_{i}$
using only the first $M\leq n_{\rm bin}$ modes (see
Eq. \ref{eq:pcaexp}). Considering all the $n_{\rm bin}$ modes ensures
that no information is thrown away. However this is achieved at the
cost that errors in the recovered quantities would be very large owing
to the presence of the negligibly small eigenvalues. On the contrary,
lowering the number of modes can reduce the error but may introduce
large biases in the recovered quantities.  An important step in this
analysis is therefore, to decide on the number of modes $M$ to be
used. In order to test this we consider a constant bias model to
  represent the true model as against the fiducial model. For a given
  data, figure \ref{fig:recon} shows how the true model is
  reconstructed through the inclusion of more and more PCA modes.  The
  reconstruction is directly related to the quality of the data. In
  the $k$-range where data is not available, the reconstruction is
  poor and the fiducial model is followed.  The reconstruction is also
  poor for large departures of the true model from the fiducial
  model. We see that a resonable reconstruction is obtained using the first $5$ modes
for $k < 1$ where the data is available.
In order to fix the value
  of $M$, we have used the Akaike information criterion
  \cite{Liddle2007} ${\rm AIC}=\chi_{\rm min}^2+2M$, whose smaller
  values are assumed to imply a more favored model.  Following the
  strategy used by \citeN{Clarkson2010} and \citeN{mitra2}, we have
  used different values of $M$ (2 to 6) for which the AIC is close to
  its minimum and amalgamated them equally at the Monte Carlo stage
  when we compute the errors. In this way, we ensure that the inherent
  bias which exists in any particular choice of $M$ is reduced.

We next perform the Monte-Carlo Markov Chain (MCMC) analysis over the
parameter space of the optimum number of PCA amplitudes $\{m_p\}$ and
$\bar{x}_{\rm HI}$.  Other cosmological parameters are held fixed to
the WMAP7 best-fit values (see Section
\ref{sec:simulation_results}). We carry out the analysis by taking
$M=2$ to $M=6$ for which the AIC criterion is satisfied. By equal
choice of weights for $M$ and folding the corresponding errors
together we reconstruct $b(k)$ and thereby $C_{\ell}$ along with their
effective errors. We have developed a code based on the publicly
available COSMOMC \cite{Lewis2002} for this purpose. A number of
distinct chains are run until the Gelman and Rubin convergence
statistics satisfies $R - 1 < 0.001$.  We have also used the
convergence diagnostic of Raftery \& Lewis to choose suitable thinning
conditions for each chain to obtain statistically independent samples.
\section{Results and Discussion}
\label{sec:discussions}

\begin{figure}
  \includegraphics[width=0.48\textwidth,height=0.34\textheight,angle=0]{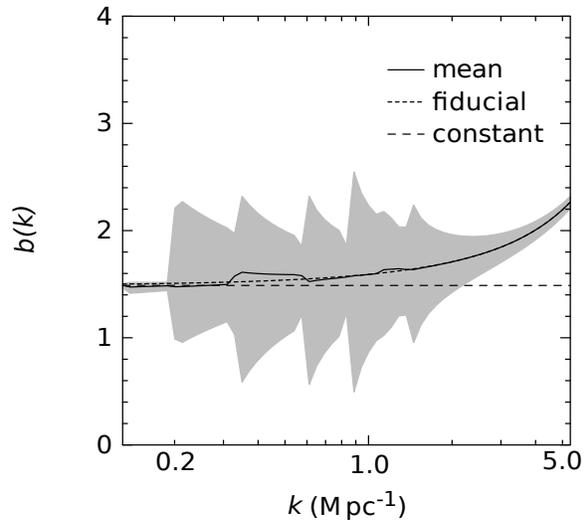} 
  \caption{The marginalized posteriori distribution of the binned bias
    function obtained from the MCMC analysis using the AIC criterion
    up to first 6 PCA eigenmodes.  The solid lines shows the mean
    values of bias parameters while the shaded regions represent the
    2-$\sigma$ confidence limits. In addition, we show the fiducial
    and constant bias models.}
\label{fig:mcmcbias}
\end{figure}

\begin{figure}
  \includegraphics[width=0.48\textwidth,height=0.34\textheight,angle=270]{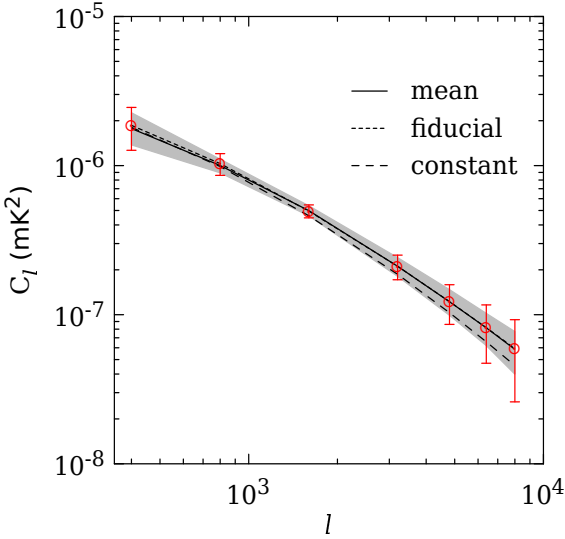}
  \caption{The reconstructed $C_{\ell}$ with its 2-$\sigma$ confidence
    limits.  The points with error-bars denote the observational
    data. The solid, short-dashed and long-dashed lines represent
    $C_{\ell}$ for the mean, fiducial and constant bias models
    respectively.}
\label{fig:mcmccl}
\end{figure}
\begin{figure}
  \includegraphics[width=0.48\textwidth,height=0.34\textheight,angle=270]{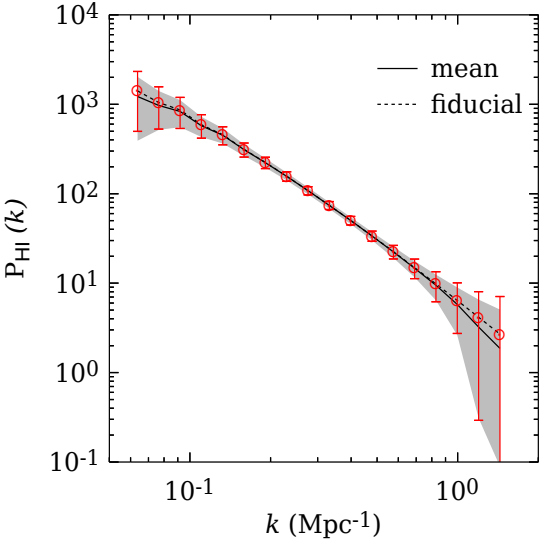}
  \caption{The reconstructed $P_{\rm HI}(k)$ with its 2-$\sigma$
    confidence limits.  The points with error-bars denote the
    observational data. We have taken $\mu = 0.5$ and $z = 2.5$}
\label{fig:mcmcpk}
\end{figure}

The reconstructed bias function obtained using the analysis described
in the last section is shown in Figure \ref{fig:mcmcbias}. The solid
line represents the mean model while the shaded region corresponds to
95\% confidence limits (2-$\sigma$).  We have also shown the fiducial
model (short-dashed) as well as the popularly used constant bias, $b
\sim 1.5$ model (long-dashed) for comparison.  We find that the
fiducial model is within the 95\% confidence limits for the entire
$k-$range considered, while the constant bias is within the same
confidence limits only up to $k\approx2$ $\rm Mpc^{-1}$.  We note that
the errors decrease drastically for $k>2$ and $k<0.2$ $\rm
Mpc^{-1}$. This is expected from the nature of the Fisher matrix which
shows that there is practically no information in the PCA modes from
these $k-$regions. Therefore, all models show a tendency to converge
towards the fiducial one. This is a direct manifestation of lack of
data points probing these scales. Thus, most of the information is
concentrated in the range $0.2 < k < 2$ $\rm Mpc^{-1}$ within which
reconstruction of the bias function is relevant with the given data
set.

The mean reconstructed bias simply follows the fiducial model for $0.2
< k < 2$ $\rm Mpc^{-1}$. This is expected as the simulated $C_{\ell}$
data is generated using the fiducial bias model itself (Section
\ref{sec:simulated_data}).  In the case of analysis using real
observed data this matching would have statistical significance,
whereas here this just serves as an internal consistency check.  The
shaded region depicting the errors around the mean is however
meaningful and tells us how well the given data can constrain the
bias. The outline of the 2-$\sigma$ confidence limits shows a jagged
feature which is directly related to the presence of the data
points. We observe that apart from the fiducial model, a constant bias
model is also consistent with the data within the 2-$\sigma$ limits.
In fact, other than imposing rough bounds $1\lesssim b(k)\lesssim2$,
the present data can hardly constrain the scale-dependence of bias. It
is also not possible for the $C_{\ell}$ data with its error-bars to
statistically distinguish between the fiducial and the constant bias
model in $0.2 < k < 2$ $\rm Mpc^{-1}$.  Figure \ref{fig:mcmccl}
illustrates the recovered angular power spectrum with its 95\%
confidence limits. Superposed on it are the original data points with
error-bars.  We also show the angular power spectrum calculated for
the fiducial and the constant bias models. The 2-$\sigma$ contour
follows the pattern of the error-bars on the data points. It is
evident that the data is largely insensitive (within its error-bars)
to the different bias models. Hence the $k-$dependence of bias on
these scales does not affect the observable quantity $C_{\ell}$ within
the bounds of statistical precision.

\begin{table}
\centering
\begin{tabular}{c|c}
\hline
\hline
Parameters & 2-$\sigma$ errors\\
\hline
$\bar{x}_{\rm HI}$ & $1.06\times10^{-3}$\\
$b_{\rm lin}$ & $0.453$\\
\hline
\hline
\end{tabular}
\caption{The 2-$\sigma$ errors for 
  $\bar{x}_{\rm HI}$ and $b_{\rm lin}(k=0.3$ $\rm Mpc^{-1})$ 
  obtained from the current analysis using AIC criterion.}
\label{tab:AIC}
\end{table}

While constructing the Fisher matrix, we had marginalized over the largely 
unknown parameter $\bar{x}_{\rm HI}$. Treating it as an independent free 
parameter, we have investigated the possibility of constraining the 
neutral fraction using the simulated $C_{\ell}$ data. 
The 2-$\sigma$ error in this parameter 
obtained from our analysis is shown in Table \ref{tab:AIC}. 
We had used the fiducial value $\bar{x}_{\rm HI} = 2.45 \times 10^{-2}$
in calculating $C_{\ell}$. It is not surprising that our analysis gives a 
mean $\bar{x}_{\rm HI} = 2.44\times 10^{-2}$ which is in excellent agreement with the fiducial value.
It is however more important to note that the given data actually constrains $\bar{x}_{\rm HI}$
reasonable well at $\sim 4 \%$.

Noting that, on large scales ($k\lesssim 0.3$ $\rm Mpc^{-1}$), one
cannot distinguish between the mean, fiducial and the constant bias
models, we use $b_{\rm lin} (= 1.496)$ to denote the bias value on
these scales. The 2-$\sigma$ error on $b_{\rm lin}$ is evaluated at
$k=0.3$ $\rm Mpc^{-1}$ (shown in Table \ref{tab:AIC}).

In the $k-$range of our interest, the fiducial model does not reflect
significant departure from the constant bias. Further, the confidence
interval obtained from the data also reflects that the observed
$C_{\ell}$ is insensitive to the form of bias function $b(k)$ in this
range - provided that it is bound between approximate cut-offs
($1\lesssim b(k)\lesssim 2$). Moreover, the bias largely affects the
amplitude of the angular power spectrum and has only a weak
contribution towards determining its shape. A scale independent
large-scale bias seems to be sufficient in modelling the data. The
mean neutral fraction which globally sets the amplitude of the power
spectrum is hence weakly degenerate with the bias. This is manifested
in the fact that though $\bar{x}_{\rm HI}$ is rather well constrained,
the bias reconstruction which uses the degraded Fisher information
(after marginalizing over $\bar{x}_{\rm HI}$) is only weakly
constrained from the same data. A prior independent knowledge about
the post reionization neutral fraction would clearly ensure a more
statistically significant bias reconstruction with smaller errors.

Figure \ref{fig:mcmcpk} shows the reconstructed 3D HI power spectrum.  The
direct algebraic relationship between the observable $P_{\rm HI}(k)$
and the bias $b(k)$ makes the 3D analysis relatively
straightforward. This is specifically evident since the Fisher matrix
elements in this case are non-zero only along the diagonal at specific
$k-$ values corresponding to the data points.  The entire routine
repeated here yields similar generic features. However, the key
difference is that we have a larger number of bins with high
sensitivity leading to an improved constraining of bias $ 1.3 < b(k) <
1.7$ in the range $ 0.2 < k < 0.7$ $\rm Mpc^{-1}$.

  In the absence of real observed data, our proposed method applied
  on a simulated data set, reflects the possibility of constraining
  large-scale $\rm HI$ bias.  The method is expected to yield better
  results if one has precise knowledge about the neutral content of
  the IGM and the underlying cosmological paradigm.  We note that the
  problem of constraining an unknown function given a known data dealt
  in this work is fairly general and several alternative methods maybe
  used.  The chief advantage of the method adopted here, apart from its
  effective data reduction, is its model independence.  The
  non-parametric nature of the analysis is specially useful in the
  absence of any specific prior information. A straightforward fitting
  of a polynomial and estimating the coeffecients may turn out to be
  effective but there is no a priori reason to believe that it would
  work.  It is logically more reasonable not to impose a model (with
  its parameters) upon the data, and instead, let the data reconstruct
  the model.

With the anticipation of upcoming radio observations towards
measurement of $\rm HI$ power spectrum, our method holds the promise
for pinning down the nature of $\rm HI$ bias thereby throwing valuable
light on our understanding of the $\rm HI$ distribution in the diffuse
IGM.

\section*{Acknowledgements}

Computational work for this study was carried out at the Centre for
Theoretical Studies, IIT, Kharagpur and the cluster computing facility
of Harish-Chandra Research
Institute\footnote{http://cluster.hri.res.in/index.html}.  Suman
Majumdar would like acknowledge Council of Scientific and Industrial
Research (CSIR), India for providing financial assistance through a
Senior Research Fellowship (File No. 9/81(1099)/10-EMR-I).  The
authors would like to thank Prof. Somnath Bharadwaj for useful
discussions and help.

\bibliography{mnrasmnemonic,IGM-ADS}
\bibliographystyle{mnras}

\end{document}